\documentclass[fleqn,10pt]{wlscirep}
\usepackage[utf8]{inputenc}
\usepackage[T1]{fontenc}
\usepackage[english]{babel}
\usepackage{graphicx}
\usepackage{subcaption}
\usepackage{siunitx}
\usepackage{placeins}
\usepackage{helvet}
\usepackage{booktabs}
\usepackage{lineno}
\title{How to Tune Autofocals: A Comparative Study of Advanced Tuning Methods}

\author[1,*]{Benedikt W. Hosp }
\author[1,2]{Yannick Sauer}
\author[1]{Björn Severitt}
\author[1]{Rajat Agarwala}
\author[1,2]{Siegfried Wahl}

\affil[1]{ZVSL, Institute for Ophthalmic Research, University of Tübingen, Maria-von-Linden Straße 6, 72076 Tübingen, Germany}
\affil[2]{Carl Zeiss Vision International GmbH, Turnstraße 27, 73430 Aalen, Germany}

\affil[*]{benedikt.hosp@uni-tuebingen.de}

\begin{abstract}
This study comprehensively evaluates tuning methods for autofocal glasses using virtual reality (VR), addressing the challenge of presbyopia. With aging, presbyopia diminishes the eye's ability to focus on nearby objects, impacting the quality of life for billions. Autofocals, employing focus-tunable lenses, dynamically adjust optical power for each fixation, promising a more natural visual experience than traditional bifocal or multifocal lenses. Our research contrasts the most common tuning methods—manual, gaze-based, and vergence—within a VR setup to mimic real-world scenarios. Utilizing the XTAL VR headset equipped with eye-tracking, the study replicated autofocal scenarios, measuring performance and usability through psychophysical tasks and NASA TLX surveys. Results show varying strengths and weaknesses across methods, with gaze control excelling in precision but not necessarily comfort and manual control providing stability and predictability. The findings guide the selection of tuning methods based on task requirements and user preferences, highlighting a balance between precision and ease of use.
\end{abstract}
\begin{document}

\flushbottom
\maketitle
%
%
\thispagestyle{empty}

\section*{Introduction}
The eye's natural aging process leads to the common condition of presbyopia, resulting in a decline in the eye's ability to focus on nearby objects \cite{atchison1995accommodation,charman2008eye}. This widespread issue significantly affects billions of individuals' visual acuity and quality of life as they age \cite{frick2015global,tahhan2013utility}. Traditional approaches, such as bifocal or multifocal lenses, impose limitations and cannot fully replicate the eye's natural accommodation \cite{lord2002multifocal,johnson2007multifocal}. These solutions restrict gaze behavior to predefined lens regions, limiting visual convenience and field of view. Furthermore, progressive lenses, while effective in correcting vision, can introduce optical aberrations that compromise visual perception \cite{charman2014developments,pope2000progressive,agarwala2023feasibility,jarosz2023pilot,sheedy2005progressive,meister2008progress,selenow2002assessing,sauer2022self}.
Autofocals are gaining attention as a new spectacle solution for presbyopia \cite{agarwala2022evaluation,wang2014pair,padmanaban2019autofocals,hasan2017adaptive}. These spectacles utilize focus-tunable lenses that dynamically adjust their optical power to focus on the currently fixated object \cite{agarwala2023feasibility}. Unlike traditional spectacles with predefined lens regions, autofocals offer the potential for a visual perception that closely resembles the natural healthy accommodation of the eye. Adjusting the focus for every new fixation aims to provide a more seamless and natural visual experience. To understand the environment and perceive the possible distances in there, different kinds of sensors, e.g., LiDAR (Light Detection and Ranging) or stereo cameras \cite{agarwala2022evaluation,padmanaban2019autofocals} have been proposed. Other research focuses on the lens design, such as Hasan et al. using liquid membrane\cite{hasan2017tunable,agarwala2022evaluation} or Li et al. \cite{li2006switchable} with liquid crystal-based lenses. Although the technology of autofocals is quite a young research field, more scientists focus on optimal hardware settings. One important aspect that has not been researched well is the user's interaction method (or tuning method) to tune the autofocals. The most common tuning methods are gaze, vergence, and manual \cite{de2021morrow}. The aim is to find a method that enables smooth and accurate focus control to reach a high usability. However, there has been no systematic comparison between different tuning methods regarding performance and convenience. As a consumer device, user experience is an essential part of the design decision of autofocals. To adequately compare the most common methods, it is necessary to recreate realistic scenarios involving dynamic changes in gaze distance. Virtual reality (VR) can provide several benefits to evaluating autofocals as it is easy to recreate a natural scene with high experimental control. This allows for precise measurement and analysis of the performance and convenience of autofocals, providing valuable insights for further development and optimization. \\
This work presents a user study that exhaustively compares different tuning methods for autofocals in virtual reality regarding performance and usability ratings. To simulate appropriate scenarios, we use  VisionaryVR \cite{visionaryvr}, a virtual reality simulation tool for optical vision correction, which offers a novel approach to evaluate and optimize autofocal algorithms and methods. The simulation tool realistically replicates optical aberrations, depth of field, and other factors affecting focus performance. It enables precise visual performance and convenience evaluation by recreating dynamic gaze distance changes using psychophysical paradigms and behavioral quantifiers. To address the limitations of current approaches of autofocal tuning methods and explore their potential, the paper evaluates a task-based evaluation framework that recreates natural scenarios involving dynamic changes in gaze distance. We aim to comprehensively assess visual performance and convenience, simulating real-life situations where individuals with presbyopia frequently experience shifts in focus. Through experiments and data analysis, the paper aims to gain insights into the effectiveness and efficiency of these control algorithms.

\section*{Methods}

The method section describes the setting of the experiment environment and the hardware and software of the data-collection procedure for reproducing the results. In the second part, we describe the data collection process and the experimental design for our user study.
\paragraph*{VR \& ET}
\label{sec:settings}
For the experiment, we utilized the XTAL VR headset, developed by VRgineers Inc, Prague, Czech Republic, in the simulation tool to replicate autofocals. The XTAL headset boasts a high-resolution display of 8k (4k for each eye at 75 Hz frame rate) and a wide field of view (150 \textdegree horizontal and 100\textdegree vertical), enhancing the visual realism and immersion of the VR experience. The headset's high-end specifications gave participants a clear and detailed view of the virtual environment. We used the integrated eye-tracking technology of the VR headset to complement the experiment to capture eye-tracking data during appropriate VR scenarios. This allowed us to record and analyze the participants' eye movements and gaze direction throughout the experiment. By evaluating the participants' gaze behavior, we could assess the performance of the gaze-based control approach for autofocals. Combining the XTAL VR headset and eye-tracking data acquisition contributed to a comprehensive and immersive evaluation of the autofocal system's performance.

\paragraph*{Autofocal Simulation}

To simulate autofocals, our VR headset incorporated a virtual tunable optical power, representing the optical power of the simulated autofocal lens. The simulated optical power could be dynamically changed within 0 to 3 diopters at two diopters per second.  This approach ensured that only objects near the distance fitting the current optical power appeared sharp. Specifically, the simulation of autofocals in VR was achieved using a simulation of spectacle lens blur, as described in the work by Sauer et al. \cite{Sauer2022c}. The calculation of defocus blur in the simulation of autofocals considered several factors. These factors included objects' distance, the tunable lens's optical power, and pupil size. Considering these factors, the strength of blur at each point in the field of view (FoV) could be determined. The blur size, or the circle of confusion, was calculated for each pixel based on the depth buffer. The rendered image was then blurred using a disk kernel with a location-dependent size, replicating the defocus blur experienced by individuals with presbyopia. Only objects near the distance fitting the current optical power appeared sharp, mimicking the focusing behavior of autofocals.

\paragraph{Tuning methods}

In this study, we evaluated two distinct tuning methods for controlling simulated tunable lenses: manual control and gaze-based control.

\paragraph{Manual Control Method}
The manual control approach in our study required users to actively select between set focus distances. This involved choosing from three distinct power settings, each corresponding to different focal lengths: near (20 cm), intermediate (1 m), and far (6 m). For this purpose, the intuitive SteamVR Knuckles controllers \cite{valve2023steamvr} were utilized, allowing users to effortlessly transition between these settings using thumb movements on the joystick. This user-directed method afforded full control over the focal distance, catering to individual preferences. A key benefit of this approach was its stability, owing to the absence of input signal noise and the non-requirement of complex sensors. However, it's important to note the difference between this simulated interaction and real-world hardware, where users typically adjust focus by physically touching the frame of the glasses, a function simplified in our setup as a button press.

\paragraph{Gaze-Based Control Method}
The gaze-based control method in our study provided an automated system for adjusting focus, driven by the user's gaze within a virtual reality environment. Utilizing sophisticated eye-tracking technology, this approach dynamically updated the focus distance based on where the user looked. The system determined the focus distance by identifying the intersection of the user's gaze with virtual objects. This method was designed to simulate a potential real-world application of autofocals, where a distance sensor, such as LIDAR or a stereo camera, could be employed. Such sensors would enhance the accuracy of the system by mapping the gaze position to a precise distance distribution. This feature would significantly benefit visual performance, aligning the focus adjustments more accurately with the user's natural gaze behavior. However, challenges such as noise, delays, and accuracy concerns in the eye-tracking system could still impact the fidelity of the focus distance estimations. This addition suggests a direction for future development in autofocal technology, enhancing its practicality and user experience.

\paragraph{Experiment}
Our data set consisted of 21 emmetropic individuals, 11 males, and 10 females, with an average age of 31.68 years (standard deviation = 11.71). All participants were free from any known eye diseases. Of these, 9 had experience with eye-tracking technologies, and 16 had been previously exposed to virtual reality environments. 
The study followed ethical guidelines and received approval from the Faculty of Medicine Ethics Committee of the University of Tübingen under reference number 439/2020BO. Written consent was collected. 
Our experiment involved a matching task developed in VisionaryVR\cite{visionaryvr}, designed to imitate daily life scenarios with dynamic eye movements. Participants were presented with stimuli on three screens at varying distances: a smartphone at 30 cm, a computer monitor at 1 m, and a television screen at 6 m. The task required participants to match a Landolt ring on one screen and a random Sloan letter on another, determining if they corresponded to the same column on a third screen (for specific implementation insights, please see \cite{visionaryvr} or see Figure \ref{fig:matching_task} which is taken from VisionaryVR \cite{visionaryvr}). This activity necessitated shifting focus across all three distances, mirroring real-life situations. The table display was randomized to ensure diverse and comparable gaze shifts, preventing a fixed screen-viewing sequence. Participants tested both tuning methods in separate sessions, completing 30 tasks per method. The stimuli, distributed normally, appeared at the screen's center or corners, incorporating complex scenarios with stimuli near the borders of different depth fields. Following each session, participants completed a NASA Task Load Index (TLX) survey in VR, providing feedback on task ease and workload. The survey dimensions included:

\begin{figure}[h]
\centering
\includegraphics[width=\textwidth]{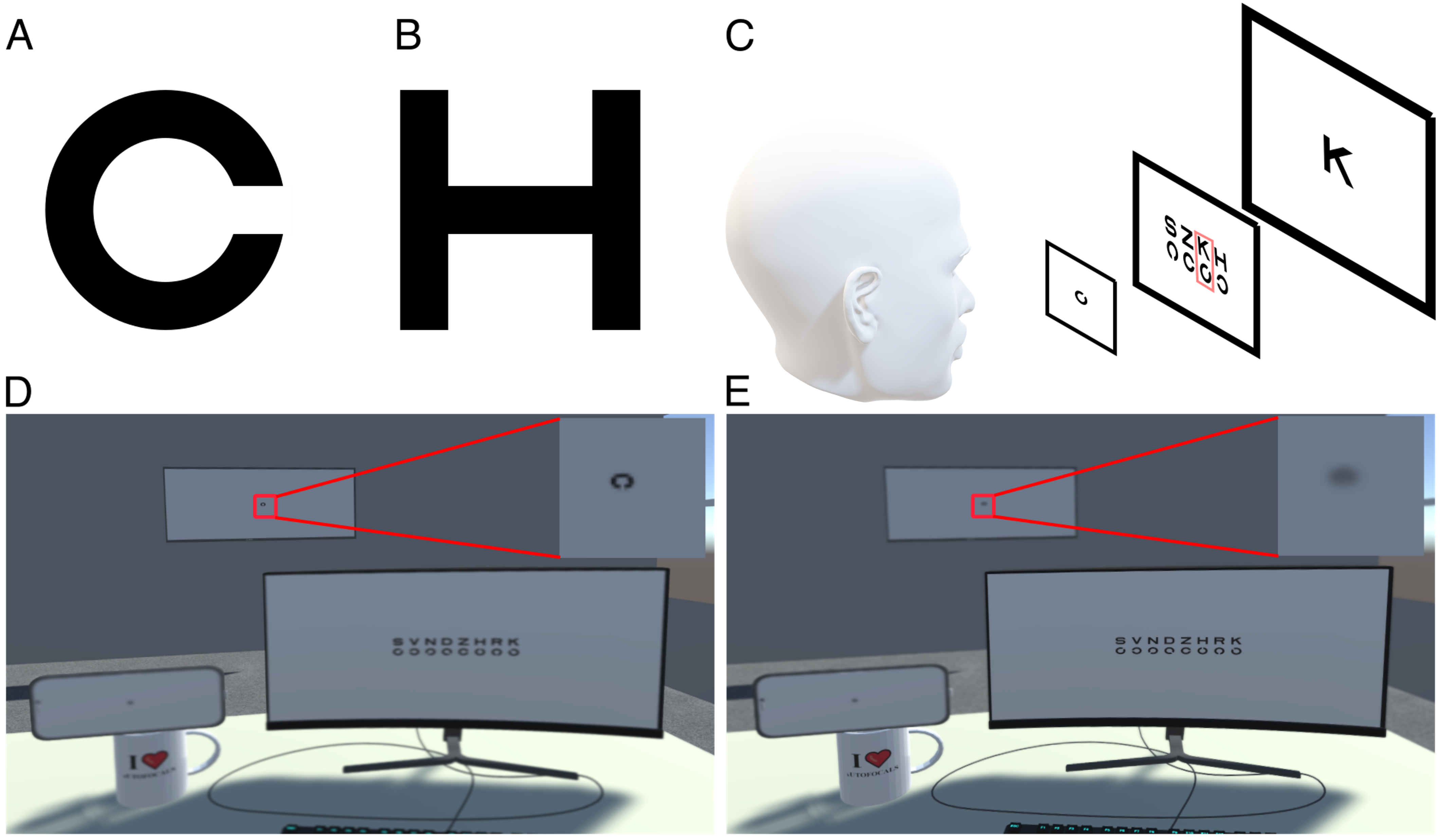}
\caption{\label{fig:matching_task} Key Elements of the Matching Task Demonstrating Focus Adjustments from VisionaryVR \cite{visionaryvr}: A) The Landolt ring, used as the initial stimulus in the task, features an opening that can face any of eight directions. B) A second screen displays one of eight Sloan letters. C) Three screens showcase stimuli at varying distances: a smartphone at \SI{30}{\cm}, mimicking reading distance; a computer display at \SI{1}{m} for intermediate vision; and a far-distance visualization at \SI{6}{m}. The task involves identifying if a Landolt ring and a Sloan letter, shown on two separate screens, appear in the same column on a third screen, which displays a table of both stimuli. This task necessitates shifting focus across all three distances. D) A virtual environment screenshot of the matching task, where defocus blur is simulated based on local depth and the adjustable focus distance of the virtual autofocal lens. Here, the focus is set to the distant screen. E) An identical scene, but with the focus adjusted to the intermediate distance, resulting in the Landolt stimulus appearing blurred and out of focus on the distant screen.}
\end{figure}

\begin{enumerate}
\item \textbf{Mental Demand}: Assessing cognitive load and mental effort required for the task. Higher scores indicate greater mental demand.
\item \textbf{Physical Demand}: Evaluating the physical effort needed for task execution. Higher scores suggest increased physical exertion.
\item \textbf{Temporal Demand}: Measuring perceived time pressure or constraints during the task. Higher scores reflect more intense time demands.
\item \textbf{Performance}: Gauging participants' self-assessment of their task performance. Higher scores denote poorer self-perceived performance.
\item \textbf{Effort}: Assessing overall mental and physical effort invested in task completion. Higher scores point to greater perceived effort.
\item \textbf{Frustration}: Determining the level of annoyance or frustration experienced during the task. Higher scores indicate increased frustration.
\end{enumerate}

\paragraph*{Experiment Procedure}
The overall experiment procedure consisted of the following steps.

\begin{itemize}

\item \textbf{Familiarization}: Subjects performed a few trials without the simulated blur of autofocals to become familiar with the task and the virtual environment.
    
\item \textbf{Explanation of Tuning Methods:} The two different tuning methods, manual control and gaze-based control, were explained to the subjects. They were informed about the operation and capabilities of each method.

\item \textbf{Randomized Trials:} Subjects operated both tuning methods in separate trials. The order of the tuning algorithms used in each trial was randomized to avoid any bias or order effects. Each subject completed 30 tasks for each tuning method, resulting in 60 tasks.

\item \textbf{Stimulus Presentation:} Stimuli were presented on three different screens at different distances. The screens represented a smartphone at a reading distance of 30 cm, a computer display at 1 m, and a far-distance TV screen at 6 m. The stimuli consisted of Landolt rings and Sloan letters, standardized optotypes for visual acuity testing.

\item \textbf{Matching Task:} The task for the subjects was to compare if the combination of the Landolt ring and Sloan letter on different screens belonged to the same column in a table displayed on a third screen. This required a focus shift between the three viewing distances and dynamic gaze changes. The screen showing the table was randomized to prevent the subjects from developing a fixed strategy for fixating the screens.

\item \textbf{NASA TLX Questionnaire:} After each trial, a NASA TLX questionnaire \cite{feick2020virtual} was presented in VR to assess the subjective task load and convenience experienced by the subjects. The NASA TLX questionnaire is used to determine subjective workload and task performance. It comprises six dimensions: Mental Demand, Physical Demand, Temporal Demand, Performance, Effort, and Frustration. Participants rate each dimension on a scale from low to high or good to poor, indicating their perceived experience in each aspect. The questionnaire provides insights into participants' subjective workload and helps evaluate the convenience and user experience of the different tuning methods for autofocals. 
\end{itemize}

The stimuli on the screens were presented at the center or the corner of the screen, with their positions randomized. This randomization incorporated complex cases where a stimulus was close to the border of two distinct depths, making the task more challenging and representative of real-world scenarios.

\section*{Evaluation}
We employed several performance metrics to compare and evaluate the performance of the tuning methods.
These metrics included Pearson correlation, root mean square error (RMSE), cosine similarity, mean absolute error (MAE), cross-correlation, and dynamic time warping (DTW). Each similarity metric contributed to the evaluation by assessing the similarity, agreement, or alignment between the control signal (the desired focus distance) and the ground truth (the target distance). The specific formulas or calculations associated with each metric were employed to quantify the performance of the tuning methods accurately.

\paragraph*{Similarity Metrics}

To evaluate the performance of the tuning methods, we employed several similarity metrics that captured different aspects of agreement and similarity between the control signal and the ground truth. The chosen similarity metrics were:

\begin{itemize}
\item \textbf{Pearson correlation:} The Pearson correlation coefficient measures the linear relationship between two variables, in this case, the control signal and the ground truth. A higher correlation coefficient indicates a stronger linear relationship and better agreement between the two signals.

\item \textbf{RMSE:} The root mean square error quantifies the average magnitude of the differences between the control signal and the ground truth. A smaller RMSE value indicates a closer agreement between the two signals.

\item \textbf{Cosine similarity:} Cosine similarity measures the cosine of the angle between two vectors in a multi-dimensional space. It ranges from -1 to 1, with 1 indicating identical signals and -1 indicating entirely dissimilar signals. A higher cosine similarity suggests a more substantial similarity or agreement between the control signal and the ground truth.

\item \textbf{Mean absolute error} The mean absolute error (MAE) measures the average dissimilarity between the control signal and the ground truth by calculating the average distance between the corresponding dimensions of the two vectors. A smaller mean absolut error suggests a closer match between the two signals.

\item \textbf{Cross-correlation:} Cross-correlation measures the similarity between two signals as a function of the displacement of one relative to the other. It indicates the strength and timing of their correspondence. Higher cross-correlation values indicate a more substantial similarity between the control signal and the ground truth.

\item \textbf{Dynamic Time Warping (DTW):} DTW is used to measure the similarity of two time series that may vary in time or speed. It aligns the time axis of the control signal with the ground truth, allowing for non-linear alignments. Smaller DTW values suggest a better alignment and similarity between the two signals.
\end{itemize}

By utilizing these similarity metrics, we could quantitatively assess the performance of the tuning methods based on different criteria and perspectives, providing a comprehensive assessment of their effectiveness and accuracy.

\section*{Results}

\subsection*{Ground Truth Approximation}
During the experiment, accurate control signals were obtained by tracking various objects. The position and orientation of the main camera in the VR setup provided information about the view's spatial origin. Additionally, specific objects' known position and depth allowed precisely determining a ground truth depth signal. Tracking the camera generated a signal to calculate the distance to objects throughout the experiment. Assuming that the object of interest was always on one of the three screens (mobile, screen, or TV), the object with the shortest Euclidean distance from its spatial center to the 3D gaze point was assumed to be the current object of interest. The correct depth was then determined based on the distance of the viewed objects center point.
The variance between the gaze-based method and ground truth distance estimations, it's important to elucidate several key aspects. Firstly, the gaze-based method relies on tracking the intersection point of the user's gaze within the virtual environment. This intersection, ideally, should align with the object identified in the ground truth data. However, discrepancies arise due to the complexity of accurately tracking the gaze in a dynamic virtual reality (VR) setting. The primary challenge lies in ensuring the gaze ray intersects with the intended object. Often, due to slight misalignments or user's gaze behavior, the intersection occurs with an adjacent or different object, leading to a variance from the ground truth measurements. Additionally, our method assumed the nearest object to the 3D gaze point as the object of interest, which might not always represent the user's actual focus, especially in a cluttered or densely populated VR scene. To mitigate these issues, future improvements could include enhanced gaze tracking algorithms and more sophisticated criteria for determining the object of interest. This would potentially align the gaze-based measurements more closely with the ground truth, thereby improving the system's overall accuracy and reliability.

Although the ground truth signal determination does not reflect the true signal of depth perception of the subjects, it can be used to measure the performance of the control signals, as they all have their limitations. The manual condition lacks flexibility in offering a wide range of focus distances, while the gaze-based control approach may have accuracy limitations in estimating gaze direction. Additionally, the eye vergence angle alone is insufficient for accurate depth estimation due to several factors, such as ambiguity in vergence changes, differences between accommodation and vergence, individual variability, and limited depth range.
Using the raw gaze signal directly to determine the ground truth is not the preferred solution in this study due to the inherent accuracy error of the eye tracker. The accuracy of eye-tracking devices can vary, and relying solely on the raw gaze signal may introduce inaccuracies in the ground truth determination. Instead, the study aims to estimate the best method for autofocal tuning despite the error sources and limitations present in each method, including the accuracy error of the eye tracker used to capture the gaze signal. By considering each method's accuracy limitations and potential errors, the study seeks to identify the most effective and reliable approach for autofocal tuning.

\paragraph{Uncertainty avoidance}
Ensuring an adequate distance between the objects of interest is crucial to prevent potential interference and misclassification caused by uncertainties in the gaze signal. When objects are too close together, inaccuracies or noise in the eye-tracking data may lead to false classifications of the current object of interest. This misclassification can introduce errors in the ground truth determination and, subsequently, impact the evaluation and comparison of different tuning methods for autofocals. By maintaining sufficient spacing between the objects of interest, the study aims to reduce the likelihood of misclassification and improve the accuracy of the ground truth signal. This approach ensures a more reliable comparison between tuning methods, as the ground truth accurately reflects the intended focus distance without significant distortions from uncertainties in the gaze signal. The appropriate distance between objects of interest is critical in obtaining a robust and accurate ground truth signal. It helps minimize the influence of uncertainties in the gaze signal, ensuring a fair evaluation of different autofocal tuning methods and yielding more meaningful and reliable results.

These errors are consistent across all tuning methods, ensuring a fair and meaningful comparison. However, the ground truth computation's potential errors and limitations can affect similarity measurements in the study. Assumptions about the object of interest and inaccuracies in eye-tracking measurements can introduce deviations from perfect similarity measurements. The simulation's limitations in replicating real-world conditions can also impact the accuracy of the ground truth signal and similarity measurements. It is important to consider these limitations when interpreting the similarity measurements and assessing the performance and accuracy of the tuning methods.

\subsection*{Performance of Tuning Methods}
In our research, we have chosen to use direct gaze tracking as our primary method for assessing gaze depth, as it offers superior reliability compared to relying solely on the eye vergence angle \cite{HOOGE20191,linton2020does}. While previous studies may have used the eye vergence angle as an indirect measure to approximate gaze depth, we believe that direct gaze tracking provides a more accurate and direct measurement of the person's focus point. However, to ensure a comprehensive assessment, we will still incorporate the evaluation of the vergence signal in our study. This combination of methods will allow us to understand gaze behavior better and provide more robust findings for our research objectives.

The performance of different tuning methods, namely manual control, gaze control, and vergence control, was evaluated within the simulation environment. The analysis considered various factors, including visual performance, response speed, and subjective convenience measures. Here are the findings and a discussion of the performance of each tuning method:

\paragraph{Visual Performance}

Quantitative results based on the descriptive statistics of the differences between the control and ground truth signals provide insights into visual performance. The manual control signal exhibited the smallest mean difference, median difference, and standard deviation compared to the gaze and vergence control signals. It also displayed a less skewed and heavy-tailed distribution. These results suggest that the manual control method aligns more closely with the ground truth signal, indicating better visual performance. The gaze control signal showed slightly larger deviations, skewness, and kurtosis than the manual control signal. While it still performed better than the vergence control signal, these findings suggest that the gaze control method introduces some deviations from the ground truth signal. The vergence control signal demonstrated the largest deviations, skewness, and kurtosis among the three methods. This indicates a less accurate and inconsistent alignment with the ground truth signal, resulting in poorer visual performance than the other two methods.

\begin{table}[ht]
\centering
\begin{tabular}{llllll}
\hline
\textbf{Method} & \textbf{Mean} & \textbf{Median} & \textbf{Std Dev} & \textbf{Skewness} & \textbf{Kurtosis} \\
\hline
Manual & 0.162 & 0.043 & 0.375 & 1.472 & 10.291 \\
Gaze & -0.917 & -0.373 & -0.629 & 3.170 & 21.360 \\
Vergence & -3.658 & -4.304 & 8.250 & 24.146 & 852.334 \\
\hline
\end{tabular}
\caption{Error between control and ground truth signal in meters.}
\label{table:statistics}
\end{table}
 
The mean difference represents the average deviation between the control and ground truth signals. In this particular analysis, the vergence control signal exhibits the highest mean difference (-3.658), followed by the gaze control signal (-0.917) and the manual control signal (0.162), as can be seen in Table \ref{table:statistics}. A lower mean difference signifies a closer alignment with the ground truth signal. Therefore, it can be observed that the manual control signal exhibits a slight deviation from the ground truth signal. Similarly, a lower median difference indicates a better alignment. In this comparative analysis, the manual control signal demonstrates the lowest median difference (0.043), followed by the gaze control signal (-0.373) and the vergence control signal (-4.304). The standard deviation measures the variability or spread of the differences between the control and ground truth signals. A smaller standard deviation implies less variability and greater consistency. Among the control signals, the manual control signal displays the slightest standard deviation (0.375), followed by the gaze control signal (-0.629) and the vergence control signal (8.250). When considering the symmetry and shape of the signals, skewness is employed to assess the asymmetry of the distribution of differences. Positive skewness suggests a longer tail on the right side, whereas negative skewness indicates a longer tail on the left side. In this analysis, the gaze control signal (skewness = 3.170) and the vergence control signal (skewness = 24.146) exhibit positive skewness, indicating a longer tail towards higher values. Conversely, the manual control signal demonstrates a slight positive skewness of 1.472. Kurtosis measures the heaviness of the tails in the distribution of differences compared to a normal distribution. Higher kurtosis values signify heavier tails. The vergence control signal displays the highest kurtosis (852.334), followed by the gaze control signal (21.360) and the manual control signal (10.291). All three control signals exhibit kurtosis values higher than a normal distribution (which has a kurtosis of 3), suggesting the presence of outliers or extreme values in the distributions.

\paragraph{Response Speed}

\begin{figure}[ht]
    \centering
    \includegraphics[width=0.8\linewidth]{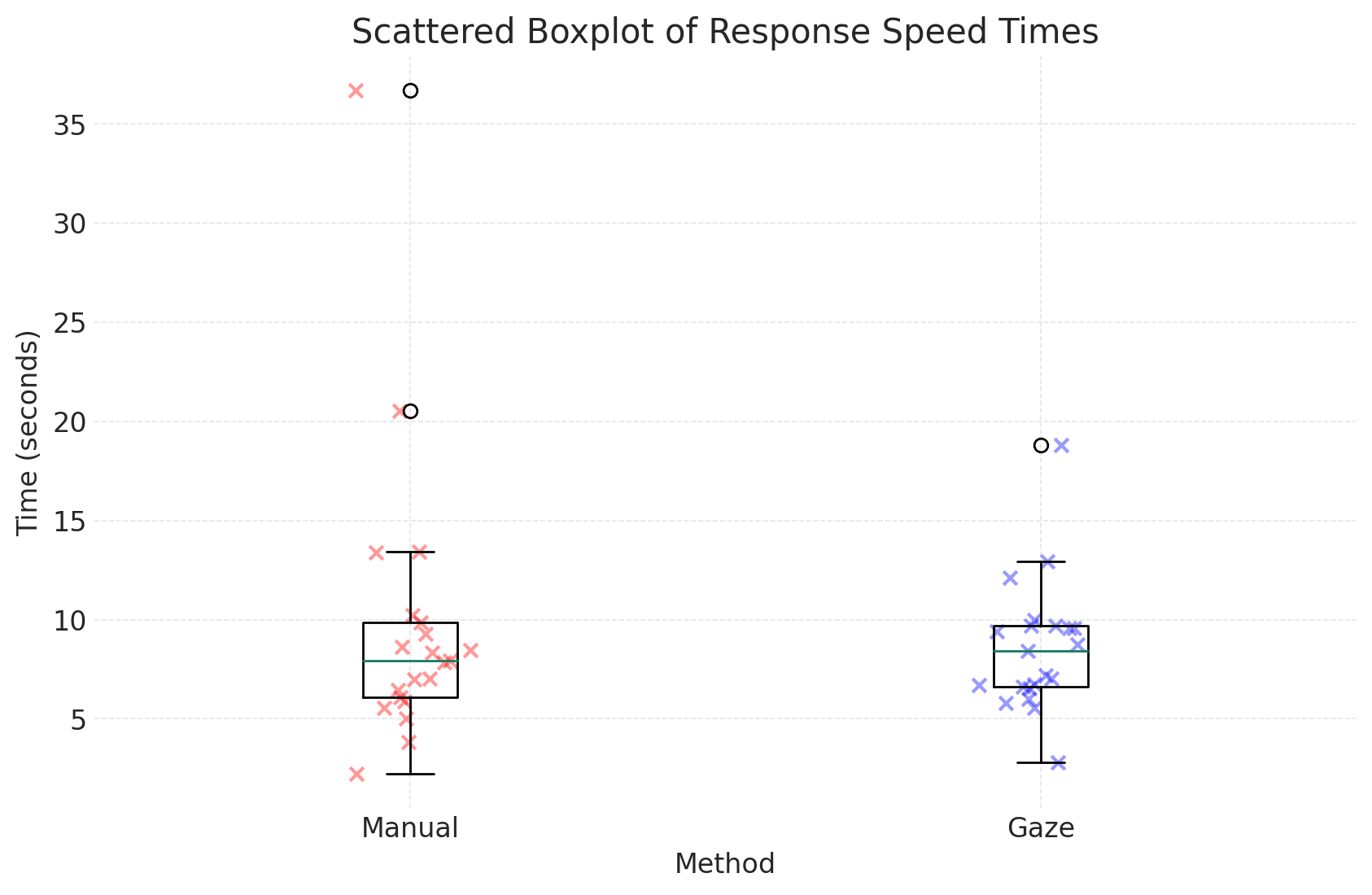}
    \caption{Comparison scattered boxplot showing non-significant differences between manual and gaze method regarding response speed.}
    \label{fig:responsespeed}
\end{figure}

The response speed of each tuning method was measured in terms of trial duration. The analysis showed that the manual control method generally had lower response times than the gaze control method. However, there were a few instances where the gaze control method exhibited faster response times. On average, the gaze control method had a slight advantage in terms of response speed. The mean response speed values indicate that, on average, the Manual control method has a slightly higher response speed (8.72 seconds) than the Gaze control method (8.57 seconds). The standard deviation values reveal that the response speed data for the Manual method has higher variability (22.00 seconds) compared to the Gaze method (4.05 seconds). This suggests that the response speed in the Manual control method varies more widely across subjects than in the Gaze control method.
Figure \ref{fig:responsespeed} provides valuable insights into the response speed performance of the Manual and Gaze control methods, highlighting differences in trial duration for different subjects and showing which method was preferred for each subject and their average performance based on response speed.

\begin{figure}[ht]
     \centering
     \includegraphics[width=0.9\linewidth]{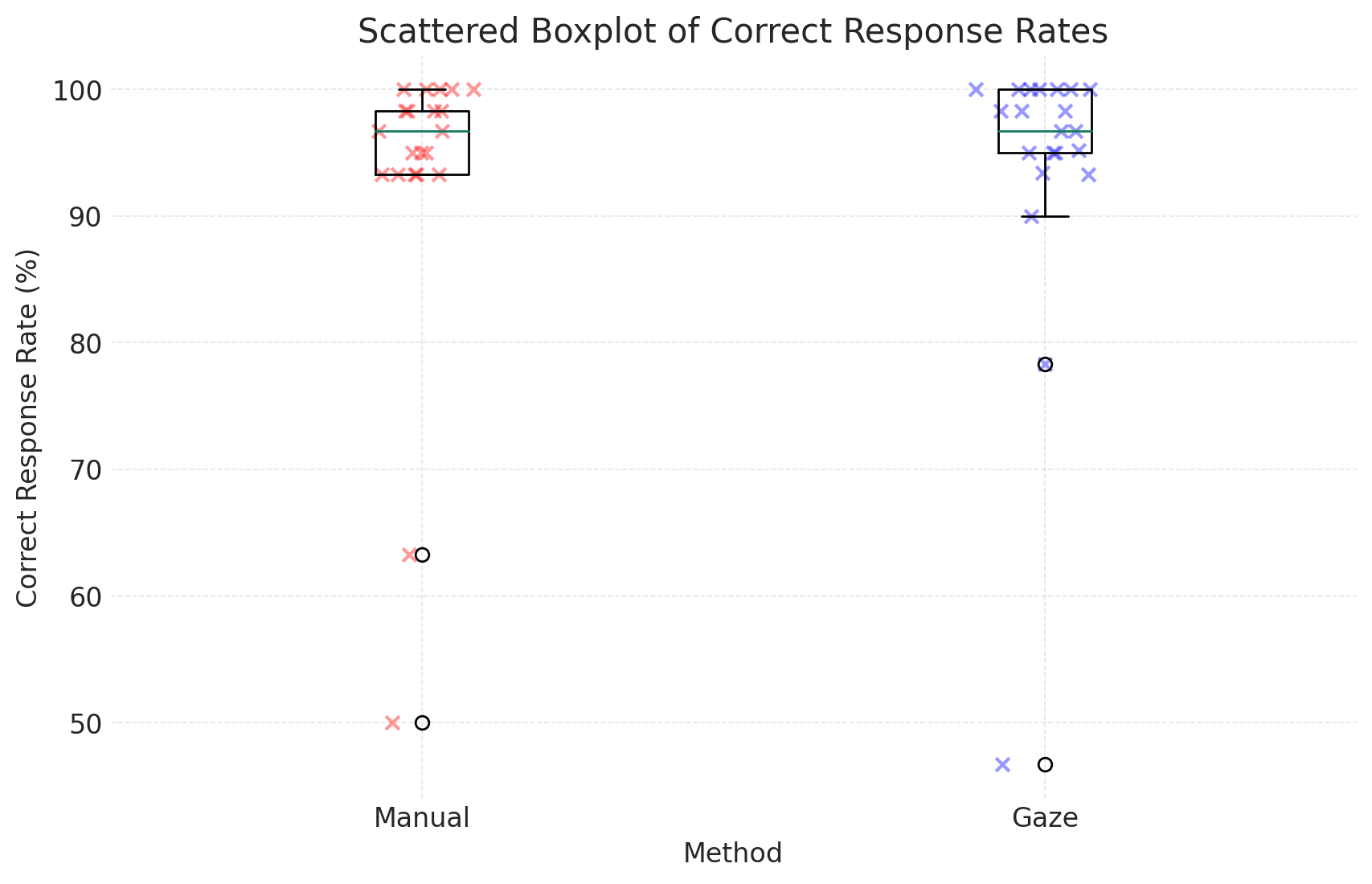}
     \caption{Comparison scattered boxplot showing the differences of the two methods in terms of correct response rate in percentage.}
     \label{fig:responserate}
 \end{figure}

\paragraph{Correct response rate}
The analysis of correct response rates across all subjects reveals that the gaze control method outperformed the manual control method in task performance. The gaze control method achieved a perfect score of 100\% in 7 out of 21 subjects, while the manual control method achieved a perfect score in 5 out of 21 subjects. The gaze control method was the preferred choice in 10 out of 21 subjects, while the manual control method won in 7 out of 21. There were four subjects where the performance between the two methods was equally good. An overview of each subject and the condition performance can be seen in Figure \ref{fig:responserate}.
Based on the correct response rate, the gaze control method exhibited a higher average performance of 94.7\%, compared to the manual control method's average performance of 94.0\%. This indicates that, on average, the manual control method resulted in slightly more accurate responses and better task performance than the gaze control method. However, the difference in performance between the two methods is relatively small and insignificant, with an average difference of only 0.71\%.
As shown in Figure \ref{fig:responserate}, the correct response rate analysis indicates that both control methods have similar task performance, with the Gaze control method having a slight advantage regarding average correct response rate and lower variability. However, other factors such as ease of use, user preference, and system requirements may also play a role in selecting the most suitable control method for specific use cases. 

    %

\begin{figure} [ht]
    \centering
    \includegraphics[width=0.9\linewidth]{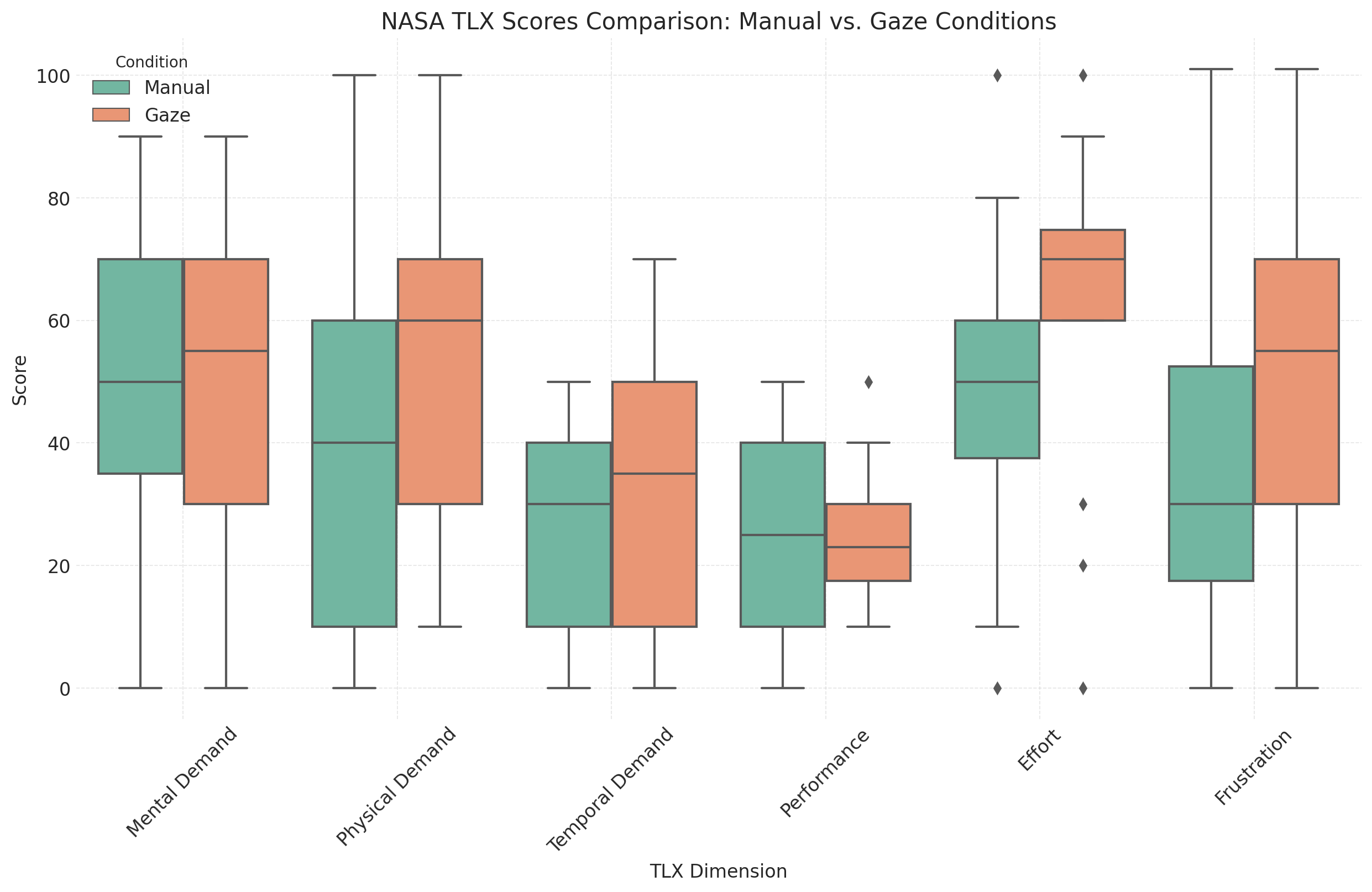}
    \caption{Overview of the NASA TLX questionnaire dimensions. The boxplots contain all data from all subjects.}
    \label{fig:nasatlx}
\end{figure}

\paragraph{Subjective Convenience Measures}
The task load evaluation included various subjective convenience measures to assess participants' experiences with each tuning method. These measures were based on the NASA Task Load Index (TLX)  ~\cite{hart1986nasa}, a widely used tool for evaluating mental workload and perceived task demands. The TLX assesses six dimensions of workload, each measured on a 60-point scale:
Participants were asked to rate each of these dimensions on a scale from 0 to 60, with 0 indicating "low" and 60 indicating "high." The scores for each dimension were then averaged to obtain a comprehensive evaluation of the subjective convenience of each tuning method. The results of the subjective convenience measures showed that the manual control method received lower scores in all dimensions (mental demand, physical demand, temporal demand, performance, effort, and frustration) compared to the gaze control method. This suggests that participants perceived the manual control method as requiring less mental and physical effort and less time pressure, leading to lower frustration levels during task execution and being more effective. These findings indicate that the gaze control method requires more mental and physical effort from participants and may induce higher frustration levels. A direct comparison of the averaged values of the particular dimensions can be seen in Figure \ref{fig:nasatlx}.

\subsection*{Similarity measures}
\begin{figure}[htbp]
    \centering
    \includegraphics[width=1\linewidth]{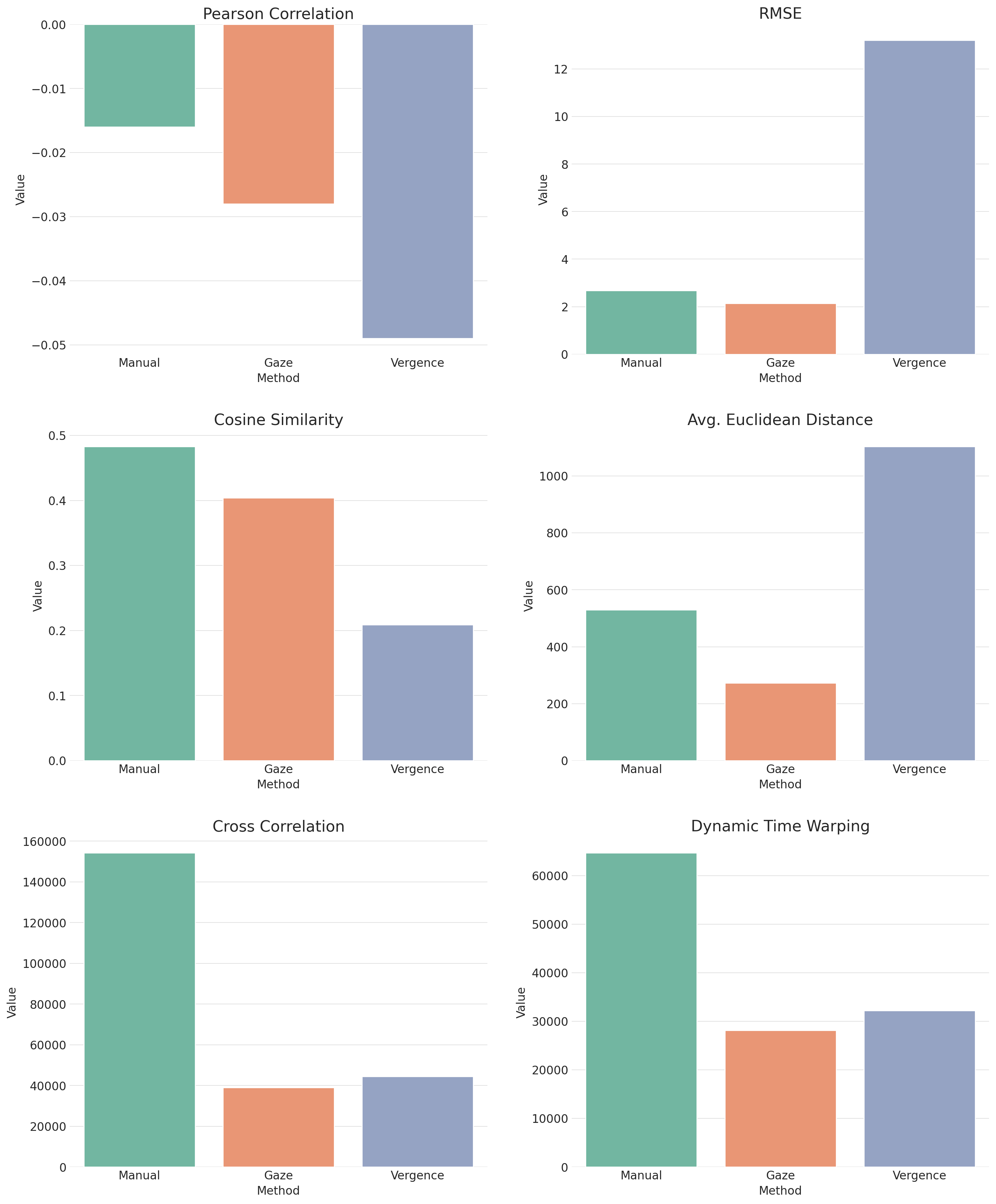}
    \caption{The bar plot shows all similarity metrics for the tuning methods next to each other. Measure are either in arbitrary units or in cm (RSME, MAE).}
    \label{fig:simmeasures}
\end{figure}

The analysis of similarity measures provides valuable insights into the performance and accuracy of the tuning methods. A comprehensive analysis of the similarity measures for the control signals, as displayed in Figure \ref{fig:simmeasures}, yields the following insightful findings. 

The Pearson correlation coefficient is a valuable tool for assessing the linear correlation between two signals, providing valuable insights into their relationship. A correlation value close to 1 indicates a strong positive correlation, while a value relative to -1 suggests a strong negative correlation. In our study, all three control signals (manual, gaze, and vergence) exhibit weak negative correlations with the ground truth signal. Specifically, the manual control signal achieves the highest correlation coefficient (-0.016), followed by the gaze control signal (-0.028) and the vergence control signal (-0.049). These relatively low correlation values indicate that the control signals poorly capture the essence or likeness of the ground truth signal. Another important measure, the root mean square error (RMSE), quantifies the average magnitude of differences between the control signal and the ground truth signal. A lower RMSE value indicates a closer alignment between the two signals. In our analysis, the gaze control signal demonstrates the lowest RMSE (2.142), followed by the manual control signal (2.681) and the vergence control signal (13.215). The significantly lower average magnitude of differences the gaze control signal exhibited suggests a relatively superior match to the ground truth signal. These values are reasonable because RMSE gives more weight to big errors as the squaring leads to larger errors disproportionately impacting RMSE. After squaring the errors, RMSE calculates their mean. This step averages out the error over all predictions, but since larger errors are squared, their impact remains significant in this average. In summary, a higher RMSE generally indicates that the signal is less accurate or has more substantial deviations from the ground truth signal. This could be due to a few very large errors or a consistent pattern of moderate errors. RMSE doesn't distinguish between these scenarios, so it's often helpful to look at other metrics or analyze error distributions to understand the differences between the two signals better. 
Cosine similarity, which measures the cosine of the angle between two vectors, assesses the similarity between the control signal and the ground truth signal. Cosine similarity values range from -1 to 1, where 1 indicates perfect similarity, and -1 represents complete dissimilarity. Our comparison reveals that the manual control signal boasts the highest cosine similarity (0.483), followed by the gaze control signal (0.404) and the vergence control signal (0.209). These values indicate that the manual control signal resembles the ground truth signal. All signals show a weak cosine similarity. Thus, this metric is not informative for our comparison. The average absolute error measures the average distance between the data points of the control signal and the ground truth signal. A lower average distance signifies a closer alignment. Within our examination, the gaze control signal demonstrates the lowest average absolute error (273.231 cm), followed by the manual control signal (529.342 cm) and the vergence control signal (1103.577 cm). The mean absolute error (MAE) treats all errors equally, regardless of their size. A higher MAE suggests that the signal consistently has larger deviations from the ground truth, but it doesn't necessarily indicate the presence of extremely large errors (like RMSE). Cross-correlation, a measure that quantifies the similarity between two signals by sliding one over the other and calculating the correlation at each step, provides further insights into their alignment. Higher cross-correlation values indicate a more pronounced similarity. In our comparison, the manual control signal exhibits the highest cross-correlation (154281.867 cm$^2$), followed by the vergence control signal (44439.620 cm$^2$) and the gaze control signal (38962.862 cm$^2$). These elevated cross-correlation values suggest a stronger similarity between the manual and vergence control signals and the ground truth signals. Dynamic time warping, which quantifies the similarity between two time series by warping the time axis to align data points, offers valuable insights into their alignment. Lower dynamic time-warping values indicate a closer alignment. In our investigation, the gaze control signal demonstrates the lowest dynamic time warping value (28140.367 cm), followed by the vergence control signal (32185.166 cm) and the manual control signal (64675.913 cm). The superior alignment of the gaze control signal with the ground truth signal is evident in these results.\\

\subsection*{Discussion}

The analysis of the autofocal system's performance reveals notable distinctions among the control signals. Gaze control notably excels in precision, as highlighted by its low RMSE and MAE values, alongside a significant cosine similarity. This suggests its strong alignment with desired outcomes, despite less impressive dynamic time-warping scores. Conversely, manual control displays superior Pearson correlation and cross-correlation, indicating a more linear and consistent relationship with the standard signal. In terms of performance, vergence control lags, evidenced by its higher RMSE, reduced cosine similarity, expanded MAE, and greater dynamic time warping, highlighting a diminished correlation with the benchmark.

Exploring deeper, gaze control shows superiority in areas like RMSE and MAE, indicating high precision. Manual control, on the other hand, exhibits better correlation metrics, aligning more closely with anticipated results. Gaze control's low RMSE and MAE suggest fine-tuned accuracy, but this doesn't necessarily equate to reduced effort, a factor that might be better evaluated through comprehensive assessments like NASA TLX.

\begin{table}[htbp]
\centering
\begin{tabular}{>{\bfseries}lccc}
\toprule
\textbf{Statistics} & \textbf{Manual} & \textbf{Gaze} & \textbf{Vergence} \\
\midrule
Response speed & - & Marginally Faster & - \\
Response rate & Comparable & Comparable & - \\
\midrule
Pearson Correlation & Most Prominent & - & - \\
RMSE & - & Most Favorable & - \\
Cosine Similarity & Most Prominent & - & - \\
MAE & - & Most Favorable & - \\
Cross-Correlation & Most Prominent & - & - \\
Dynamic Warping & - & Most Favorable & - \\
\midrule
NASA TLX & Optimal & - & - \\
\bottomrule
\end{tabular}
\caption{Comparative Overview of Autofocal Glasses Tuning Methods}
\label{tab:summary}
\end{table}

The manual approach, though slower, offers desirable control and predictability. Its linear consistency and high cosine similarity might contribute to an intuitive and user-friendly operation, albeit with slower response times. This suggests a gradual, comfortable learning curve for users. Gaze control's lower correlation, despite its precision, might imply a minor deviation from expected paths. Conversely, manual control's higher correlation, despite larger errors, shows adherence to general trends, albeit with variability, potentially influenced by fixed-depth plane selection. These insights highlight that control method selection should be based on task demands and user preferences, weighing precision against user-friendliness. The findings also point to vergence control's limitations, necessitating improvements for better performance. The study further suggests enhancing the experimental setup, particularly in gaze tracking and manual control depth step calibration. Practically, gaze control suits tasks needing quick, frequent focus adjustments where accuracy is critical, though it may increase cognitive load. In contrast, manual control is better for scenarios requiring steady, user-led operation, providing stability and comfort, particularly in less hurried environments. Ultimately, manual and gaze-based methods show comparable performance, but manual control appears to have an edge in subjective assessments. The preference for automated methods, despite associated frustrations, indicates a demand for such controls. However, limitations, particularly in eye tracking, often lead to user frustration. Addressing these issues through advanced algorithms could enhance user experience and reduce frustration.
\bibliography{main}

\section*{Acknowledgements}

The German Research Foundation (DFG) generously supported this research under SFB 1233, Robust Vision: Inference Principles and Neural Mechanisms, TP TRA, with project number 276693517.

\section*{Author contributions statement}
BWH contributed to the development, data collection, analysis, writing, and hypothesis formulation. YS contributed to the development, writing, and hypothesis formulation. RA contributed to writing and hypothesis formulation. SW contributed to writing, hypothesis formulation, and project funding. The authors declare no competing interests.


\end{document}


\appendix

\section{Questionnaire Evaluation: Manual}
\begin{table}[htbp]
    \centering
    \begin{tabular}{ccccccc}
        \toprule
        Subject & Mental Demand & Physical Demand & Temporal Demand & Performance & Effort & Frustration \\
        \midrule
        1 & 0 & 0 & 10 & 20 & 60 & 0 \\
        2 & 20 & 10 & 10 & 10 & 40 & 70 \\
        3 & 70 & 60 & 50 & 20 & 60 & 60 \\
        4 & 90 & 90 & 0 & 30 & 80 & 30 \\
        5 & 40 & 0 & 40 & 20 & 50 & 20 \\
        6 & 20 & 10 & 0 & 30 & 0 & 0 \\
        7 & 10 & 10 & 20 & 10 & 10 & 10 \\
        8 & 40 & 30 & 10 & 10 & 40 & 20 \\
        9 & 40 & 30 & 30 & 50 & 50 & 50 \\
        10 & 80 & 80 & 40 & 40 & 70 & 40 \\
        11 & 50 & 100 & 0 & 40 & 100 & 101 \\
        12 & 50 & 50 & 30 & 10 & 30 & 30 \\
        13 & 60 & 50 & 40 & 30 & 40 & 60 \\
        14 & 90 & 60 & 10 & 0 & 10 & 10 \\
        15 & 60 & 30 & 40 & 10 & 60 & 40 \\
        16 & 45 & 50 & 50 & 50 & 50 & 50 \\
        17 & 20 & 30 & 50 & 10 & 60 & 20 \\
        18 & 70 & 10 & 10 & 30 & 30 & 10 \\
        19 & 70 & 70 & 40 & 40 & 60 & 30 \\
        20 & 60 & 60 & 40 & 40 & 60 & 60 \\
        \bottomrule
    \end{tabular}
    \caption{The table shows all subjects with their rating on the NASA TLX questionnaire for the manual condition. The single columns describe different dimensions of cognitive demand. Answer rating ranges from 0 = not agree/low demand to 100 = agree/ high demand.}
    \label{tbl:manual_all_questions_all_subjects}
\end{table}

\section{Questionnaire Evaluation: Gaze}
\begin{table}[htbp]
    \centering
    \begin{tabular}{ccccccc}
        \toprule
        Subject & Mental Demand & Physical Demand & Temporal Demand & Performance & Effort & Frustration \\
        \midrule
        1 & 0 & 60 & 10 & 30 & 80 & 60 \\
        2 & 40 & 60 & 10 & 30 & 70 & 80 \\
        3 & 70 & 70 & 50 & 50 & 60 & 60 \\
        4 & 0 & 100 & 0 & 10 & 60 & 80 \\
        5 & 40 & 10 & 40 & 30 & 80 & 70 \\
        6 & 10 & 10 & 0 & 10 & 0 & 0 \\
        7 & 30 & 30 & 40 & 20 & 80 & 20 \\
        8 & 90 & 90 & 10 & 40 & 90 & 90 \\
        9 & 40 & 60 & 20 & 20 & 70 & 30 \\
        10 & 80 & 60 & 30 & 20 & 60 & 60 \\
        11 & 30 & 30 & 50 & 10 & 60 & 70 \\
        12 & 70 & 100 & 0 & 30 & 100 & 101 \\
        13 & 70 & 70 & 50 & 20 & 70 & 50 \\
        14 & 70 & 40 & 40 & 40 & 60 & 70 \\
        15 & 70 & 60 & 10 & 10 & 20 & 20 \\
        16 & 70 & 60 & 50 & 20 & 70 & 50 \\
        17 & 64 & 29 & 51 & 26 & 73 & 50 \\
        18 & 20 & 80 & 70 & 10 & 60 & 30 \\
        19 & 50 & 20 & 10 & 30 & 30 & 30 \\
        20 & 60 & 60 & 40 & 40 & 70 & 30 \\
        \bottomrule
    \end{tabular}
    \caption{Similar to table \ref{tbl:manual_all_questions_all_subjects}, the table shows all subjects with their rating on the NASA TLX questionnaire for the gaze condition. The single columns describe different dimensions of cognitive demand. Answer rating ranges from 0 = not agree/low demand to 100 = agree/ high demand.}
    \label{tbl:gaze_all_questions_all_subjects}
\end{table}